# COMMENTS ABOUT THE BOUNDARY CONDITION FOR REDUCED RADIAL WAVE FUNCTION IN MULTI-DIMENSIONAL SCHRODINGER EQUATION

ANZOR KHELASHVILI

*Inst.of High Energy Physics,,Iv.Javakhihvili Tbilisi State University*
*G.Danelia Str. 10 , Tbilisi 0109, Georgia*
anzor.khelashvili@tsu.ge

TEIMURAZ NADAREISHVILI

*Inst.of High Energy Physics,,Iv.Javakhihvili Tbilisi State University*
*G.Danelia Str. 10 , Tbilisi 0109, Georgia*
*Faculty of Exact and Natural Sciences, Iv.Javakhihvili Tbilisi State University,*
*Chavchavadze Ave 3, Tbilisi 0179,Georgia*
teimuraz.nadareishvili@tsu.ge

The problem of boundary behavior at the origin of coordinates is discussed for D-dimensional Schrodinger equation in the framework of hyperspherical formalism, which have been often considered last time. We show that the naive (Dirichlet) condition, which seems as natural, is not mathematically well justified, on the contrary to the 3-dimensional case**.** The stronger argument in favor of Dirichlet boundary condition is the requirement of time independence of wave function's norm. The problem remains open for singular potentials.

## 1. Introduction

During the last decades, number of papers have appeared about the Schrodinger equation in multi D-dimensional spaces [1-6]. For reduction of such a problem, usage of the hyperspherical formalism is the most expedient. It is noteworthy that besides the mathematical interest, this formalism has been successfully applied to various realistic physical problems, such as many particles problem, when $N$ particles are placed in $D = 3N - 1$ dimensional Euclidian space [7,8]. By its meaning $D$ dimensions has some important peculiarities in comparison to 3-dimensions. Here the central symmetry is meant using some collective potential, which has a symmetry concerning rotations relative to full $D$ - dimensional space. Owing to that in the hyperspherical basis the separation of variables takes place and the one dimensional Schrodinger equation is derived with respect to a hyperradial variable leaving some traces of hyperspherical angles.

Multidimensional central potentials with a specific analytical form have been used to interpret a number of physical phenomena and chemical processes, to explain the behavior of nanotechnological systems, etc.

The objective of the present manuscript is to focus on special features of quantum mechanics in higher dimensions, specifically on the behavior of the radial wave function at the origin of coordinates. As well as we are dealing with a second order differential equation, to impose the suitable boundary conditions are necessary for determination the structure of spectra. This problem will be considered below after introducing main equations.

## II. Preliminaries

In arbitrary D-dimensions with hypercentral potential $V_D(r)$ the Schrodinger equation has the form [1] ($\hbar = m = 1$ units are chosen)

$$\left[ -\frac{1}{2}\nabla_D^2 + V_D(r) \right] \psi_D(\mathbf{r}) = E_D \psi_D(\mathbf{r}) \qquad (1)$$

where $\mathbf{r}$ is a D-dimensional position vector, whose hyperspherical coordinates are $(r, \theta_1, \theta_2, ... \theta_{D-1})$ and $r$ is a hyperradius, denoting the radial distance

$$r = \left( \sum_{i=1}^{D} x_i^2 \right)^{1/2}. \qquad (2)$$

By $V_D(r)$ we mean a "collective" (but not a pair) potential, which depends only on $r$. The Laplacian of this problem is expressed as

$$\nabla_D^2 = \frac{1}{r^{D-1}} \frac{\partial}{\partial r} r^{D-1} \frac{\partial}{\partial r} - \frac{\Lambda_{D-1}^2}{r^2}, \qquad (3)$$

where $\Lambda_{D-1}$ is a generalized squared angular momentum in $D$ - dimensions. It obeys to the following equation

$$\Lambda^2_{D-1}\mathcal{Y}_{l,\{\mu\}}(\Omega_{D-1}) = l(l+D-2)\mathcal{Y}_{l,\{\mu\}}(\Omega_{D-1}) \qquad (4)$$

Here $\mathcal{Y}_{l,\{\mu\}}(\Omega_{D-1})$ describes hyper spherical harmonics, which are characterized by quantum numbers $(l \equiv \mu_1, \mu_2, \mu_3, ..., \mu_{D-1} \equiv m) \equiv (l,\{\mu\})$. They are natural numbers $l = 0,1,2,...;$ $l \equiv \mu_1 \geq \mu_2 \geq \mu_3 \geq ... \geq \mu_{D-2} \geq |\mu_{D-1}| \equiv m$ and these objects obey to orthogonality conditions

$$\int_{S_{D-1}} d\Omega_{D-1}\mathcal{Y}^*_{l',\{m'\}}(\Omega_{D-1})\mathcal{Y}_{l,\{\mu\}}(\Omega_{D-1}) = \delta_{ll'}\delta_{\{\mu\},\{\mu'\}} \qquad (5)$$

Substituting

$$\psi_{E,l,\{\mu\}}(\mathbf{r}) = R_{El}(r)\mathcal{Y}_{l,\{\mu\}}(\Omega_{D-1}) \qquad (6)$$

and accounting for (4)-(5), we obtain a hyperradial equation

$$\left[-\frac{1}{2}\frac{d^2}{dr^2} - \frac{D-1}{2r}\frac{d}{dr} + \frac{l(l+D-2)}{2r^2} + V_D(r)\right]R_{El}(r) = E_D R_{El}(r) \qquad (7)$$

It is a usual practice to withdraw a first derivative term, which can be achieved by substitution

$$u_{El}(r) = r^{(D-1)/2} R_{El}(r) \qquad (8)$$

After this, the equation reduces to the form

$$\left[-\frac{d^2}{dr^2} + V_{eff}(r)\right]u_{El}(r) = E_D(r)u(r) \qquad (9)$$

where

$$V_{eff}(r) = V_D(r) + \frac{L(L+1)}{r^2} = V_D(r) + \frac{l(l+D-2)}{r^2} + \frac{(D-1)(D-3)}{4r^2} \qquad (10)$$

and $L$ is a "Grand orbital quantum number"

$$L = l + \frac{D-3}{2} \qquad (11)$$

Therefore

$$l(l+D-2) = L(L+1) - (D-1)(D-3)/4 \qquad (12)$$

As it is mentioned in literature, *the physical solutions require* that $u_{El}(r) \to 0$, when $r \to 0$. Moreover, the normalization to unity of the total wave function leads to the following property of the reduced radial wave function $u_{El}(r)$:

$$\int_0^\infty u_{El}^2(r) dr = 1 \tag{13}$$

Our aim is to establish under which physical conditions follows the above mentioned boundary behavior at the origin of coordinates. We see from Eqs. (9) and (10) that the Schrodinger equation describes the one-dimensional non-relativistic motion of the particle. It is worth noting that the D-dimensional Schrodinger equation is formally the same as the radial equation in three-dimensional case, but with the grand orbital momentum $L$. The particle is subjected to the natural force coming from the potential $V_D(r)$ and two additional forces (10) with different physical origin: the centrifugal force associated with a nonvanishing hyperangular momentum, and a quantum fictitious force, associated to the quantum-centrifugal potential $(D-1)(D-3)/4r^2$ of purely dimensional origin. If we rewrite eq. (10) in more transparent form as:

$$V_{eff}(r) = V_D(r) + \frac{(D+2l)^2 - 4(D+2l) + 3}{8r^2} \tag{14}$$

we see that the effective potential depends on $D$ and $l$ through a special combination $(D+2l)$. Therefore, it appears that there is the interdimensional-degeneracy phenomenon; this implies e.g., that for an arbitrary potential the energies of the 7-dimensional s-states are the same as those of the 5-dimensional p-states or the 3-dimensional d-states.

### III. Wave function's behavior at the origin of coordinates

Let us mention that in 90-th of the previous century the problem of self-adjointness of the reduced radial Hamiltonian was a subject of intensive considerations [9]. It is well known that the self-adjointness by itself is connected to the behavior of reduced wave function at the origin. Various possibilities (choice) of boundary conditions were considered, but the final agreement was not established, especially for singular potentials. Among them was the above zero asymptotic (Dirichlet), but the problem of s-wave remained open. The so-called Robin boundary condition became preferable. We have previously proved [10] that the Dirichlet boundary condition appears to be unique. The reason is the delta-like singularity, which appears in Laplacian in the course of transition from total to reduced wave function. It is interesting if there is a such mechanism in multi dimensions.

   Note that the multi-dimensional equations reduce to the 3-dimensional ones after substitution $D = 3$. In this case, $L \to l$ and we have for total radial function the equation:

$$\left[-\frac{1}{2}\frac{d^2}{dr^2}-\frac{1}{2r}\frac{d}{dr}+\frac{l(l+1)}{2r^2}+V_3(r)\right]R_{El}(r)=E_3R_{El}(r), \quad (15)$$

while the transformation (6) reads as: $u_{El}(r)=rR_{El}(r)$ or $R=u/r$.
After this substitution it follows that the radial part of Laplacian acts on the factor $1/r$, which is a three dimensional delta function. Therefore, the additional term $\delta(r)u(r)$ appears in the reduced equation, which at this step looks like

$$\frac{r}{2}\left(-\frac{d^2u(r)}{dr^2}+\frac{l(l+1)}{r^2}u(r)\right)+\delta(r)u(r)-2m\left[E-V_3(r)\right]ru(r)=0 \quad (16)$$

The only way to avoid this extra term is a constraint $u(0)=0$. Only after using of this constraint, which has a form of Dirichlet boundary condition, we return to the generally accepted reduced equation. Moreover, this fact is valid irrespective whether the potential is regular or singular and the problem of self-adjointness of reduced Hamiltonian is also solved.

In regards to multidimensional case D>3, analogous phenomenon does not occur, because in D- dimensions delta function appears in the following equation [11]

$$\Delta_r\frac{1}{r^{D-2}}=-(D-2)\Omega_D\delta(r), \quad \Omega_D=\frac{2\pi^{D/3}}{\Gamma\left(\frac{D}{2}\right)} \quad (17)$$

and the substitution (6) has nothing in common with this equation, except for $D=3$.

Therefore, the following question arises: Is it possible, with the help of some regular steps to derive physically acceptable boundary condition? To address this problem let us draw other physical suggestions, considered e.g., in [10], which can also be transferred to D>3 spaces.

Usually, the normalization condition is considered (13) and tried to find maximal singular behavior, consistent with this condition and with the fundamental principles of quantum mechanics. Let us consider some more common physical reasonings:

From the continuity of $R(r)$ at the origin, according to (8) it follows $u(0)=0$, insuring a finite probability at this point. Exactly this idea is used in any textbook on quantum mechanics. But it is desirable to weaken this requirement, because it is too strong. One can require a finite differential probability in the spherical slice $(r, r+dr)$, or

$$|R|^2 r^{D-1}dr<\infty \quad (18)$$

If $R\sim r^s$ at the origin, we must require $2s+D-1>0$ and it follows that $s>(1-D)/2$, or

$$u_{El}(r)=r^{(D-1)/2}R_{El}(r)=r^{\frac{D-1}{2}+s}\rightarrow u(0)=0 \quad (19)$$

Another generalization is to require a finite total probability inside a sphere of small radius $a$,

$$\int_0^a |R|^2 \, r^{D-1} dr < \infty \qquad (20)$$

In this case more singular behavior is permissible, namely

$$\lim_{r \to 0} R(r) \approx \lim_{a \to 0} a^{-D/2+\varepsilon} \qquad (21)$$

where $\varepsilon > 0$ is a small positive constant and $\varepsilon \to 0$ at the end of the calculation. In this case

$$\lim_{r \to 0} u(r) \approx \lim_{a \to 0} a^{(D-1)/2} a^{-D/2+\varepsilon} \approx \lim_{a \to 0} a^{-1/2+\varepsilon} \to \infty \qquad (22)$$

The same constraint follows from the finite behavior of the norm

$$\int_0^\infty |R(r)|^2 \, r^{D-1} dr < \infty \qquad (23)$$

The strongest is the Pauli argument [12], namely, the time independence of the norm or conservation of the numbers of particles. To explore it, we follow the procedure described in [10]: In quantum mechanics the norm of the wave function is to be independent of time

$$\frac{d}{dt} \int \psi^* \psi \, dV = 0 \qquad (24)$$

By using the time dependent Schrodinger equation, we transform this equation to

$$-\frac{i}{\hbar} \int \left[ \psi^* (H\psi) - (H\psi^*) \psi \right] dV = 0 \qquad (25)$$

Thus, the time independence of probability means that the Hamiltonian must be a Hermitian operator. By introducing the probability current density

$$\mathbf{J} = \mathrm{Re}\left[ \psi^* \frac{\hbar}{im} \nabla \psi \right] \qquad (26)$$

it is easy to show that

$$div \mathbf{J} = \frac{i}{\hbar} \left[ \psi^* (H\psi) - (H\psi^*) \psi \right] \qquad (27)$$

The equation for conservation of probability takes the form (after using the Gauss' theorem)

$$\frac{d}{dt} \int_V \psi^* \psi \, dV = -\int_V div \mathbf{J} \, dV = -\int_S J_N \, dS \qquad (28)$$

where $J_N$ is the normal component of the current relative to the surface.

If we assume that at the origin the Hamiltonian has a singular point, Gauss' theorem in the last equation is not applicable. We must exclude this point from the integration volume and surround it by a small sphere of radius $a$. In this case, the surface integral is divided into a surface at infinity that encloses the total volume, and the surface of a sphere of radius $a$:

$$\lim_{a \to 0} a^{D-1} \int J_a d\Omega + \int_\infty J_N dS = 0 \qquad (29)$$

where $d\Omega$ is an element of solid angle. In the $D$-dimensions we should have $dS = a^{D-1} d\Omega$. Because the wave function must vanish at infinity, the second term goes to zero. If we substitute

$$J_a = \frac{i\hbar}{2m}\left(\left[\psi\left(\frac{\partial \psi^*}{\partial r}\right) - \psi^*\left(\frac{\partial \psi}{\partial r}\right)\right]\right)_{r=a} \qquad (30)$$

and assume $\psi = \tilde{u}/r^s$, where $\tilde{u}$ is regular at $r \to 0$, we obtain

$$\lim_{a \to 0} \frac{a^{D-1}}{a^{2s}} \int \left(\tilde{u}\frac{\partial \tilde{u}^*}{\partial r} - \tilde{u}^* \frac{\partial \tilde{u}}{\partial r}\right)_{r=a} d\Omega = 0 \qquad (31)$$

This equation is satisfied if $s < (D-1)/2$. It follows that $R(r)$ does not diverge more rapidly than $1/r^s$, but now $s < (D-1)/2$, which means that

$$\lim_{r \to 0} u(r) \approx \lim_{r \to 0} r^{-s+\frac{D-1}{2}} \to 0 \qquad (32)$$

Consequently, we see that different physically acceptable arguments lead to diverse conclusions for the wave function behavior at the origin. Namely, a finite norm allows for a certain divergent behavior of $u(r)$, but the time independence of the norm gives vanishing behavior. We are inclined to think that this last requirement is most fundamental and the vanishing of the reduced wave function is accepted as valid.

In this context, one can remember the opinion of W. Pauli [12], from his "General Principles of Quantum Mechanics", (p.45), that "An eigenfunction for which $\lim_{r \to 0}(rR) \neq 0$, is not admissible", though for such function

$$\int_0^\infty |R|^2 \, r^2 dr \qquad (33)$$

exists.

**IV. Singular potential and the self-adjoint extension (SAE)**

The behavior of reduced wave function, when $r$ turns to the origin of coordinates evidently depends on potential $V_D(r)$ under consideration. The authors of [1-6] believe that "the physical solutions require that $u_{El}(r) \to 0$ when $r \to 0$ and $\infty$". But this opinion may not be correct in general without considering singular properties of the outcome potential. It must be clarified, which physical solutions are meant by authors.

From this point of view the following classification is known for the Schrodinger equation [13] (It is natural, that this classification is the same in any dimensions):

- (1) **Regular potentials**. They behave as

$$\lim_{r \to 0} r^2 V_D(r) = 0 \tag{34}$$

For which solution of the equation

$$\left( -\frac{d^2}{dr^2} + \frac{L(L+1)}{r^2} + V_D(r) \right) u_D(r) = E_D u_D(r) \tag{35}$$

at the origin behaves like

$$u(r) \underset{r \to 0}{=} C_1 r^{L+1} + C_2 r^{-L} \tag{36}$$

Because $L$ always is positive when $D \geq 4$, the second solution must be discarded, i.e., $C_2 = 0$.

Therefore, all singularities may be contained into $V_D(r)$.

- (2) **Singular potentials,** for which

$$\lim_{r \to 0} r^2 V_D(r) = \pm \infty \tag{37}$$

For them, the "falling to the center" happens and is not interesting for us now.

- (3) **"Soft-singular" potentials,** for which

$$\lim_{r \to 0} r^2 V_D(r) = \pm V_0, \qquad (V_0 = const > 0) \tag{38}$$

Here the $(+)$ sign corresponds to repulsion, while the $(-)$ sign – to attraction.

For such potentials, the wave function has the following behavior $\lim_{r \to 0} u(r) = A_1 r^{1/2+P} + A_2 r^{1/2-P} = u_{st} + u_{add}$, where

$$P = \sqrt{(L+1/2)^2 - 2V_0} \tag{39}$$

In the region $0 < P < 1/2$ the second solution (33) satisfies also Dirichlet boundary condition and hence it must be retained in general and therefore the self-adjoint extension need to be performed [14]. As for the region $P \geq 1/2$, only the first (standard or regular) solution remains. Recalling the relation (9), one can rewrite $P$ as follows

$$P = \sqrt{\left[ l + \frac{D-2}{2} \right]^2 - 2V_0}, \tag{40}$$

or existence of the second (additional) solution can take place when

$$\left[l+\frac{D-2}{2}\right]^2 - 1/4 < 2V_0 \qquad (41)$$

i.e. with growth of dimension the restriction on $V_0$ increases. Therefore, the appearance of extra (so-called, hydrino) states becomes more limited [14].

**V. Conclusions**

In this article, we consider the problem of boundary condition of the radial wave function in an arbitrary dimensional quantum mechanics for central potentials. We have shown that in many ($D>3$) -dimensions there are no rigorous reasonings to fix boundary condition at the origin of coordinates, contrary to 3-dimensions. But from the time independence of the norm (which means a conservation of particle number in nonrelativistic quantum mechanics), the vanishing (i.e., Dirichlet) boundary condition is strongly motivated for both regular as well as singular potentials. In this respect, remarkably enough, 3 – dimensions stand out sharply against the other dimensions in the sense that only in $D=3$ – dimensions the reducing procedure automatically gives the boundary condition, $u(0)=0$ and moreover, corresponding Hamiltonian becomes a self-adjoint operator (For details, see [14]).